\documentclass[runningheads,a4paper]{llncs}
\usepackage{subcaption}
\usepackage{tikz,pgfplots}
\usepackage{amssymb}
\usepackage{graphicx}
\usepackage{url}
\usepackage{enumitem}
\usepackage{fixltx2e}
\newcommand{\subscript}[2]{$#1 _ #2$}
\newcommand{\keywords}[1]{\par\addvspace\baselineskip
\noindent\keywordname\enspace\ignorespaces#1}
\begin{document}
\mainmatter  


\title{Automatic estimation of harmonic tension by distributed representation of chords}

\titlerunning{Automatic estimation of harmonic tension}

%
%
\author{Ali Nikrang\inst{1}\and David R. W. Sears\inst{2} \and Gerhard Widmer\inst{2} \thanks{This research is supported by the European Research Council (ERC) under the EUs Horizon 2020 Framework Programme (ERC Grant Agreement number 670035, project ``Con Espressione'').}}
%
\authorrunning{Nikrang A., Sears D. Widmer, G. }
\institute{Ars Electronica Linz GmbH \& Co KG, Linz, Austria \and Johannes Kepler University, Linz, Austria\\  \email{an@musicresearch.eu}}
%
%
\maketitle
\begin{abstract}
The buildup and release of a sense of tension is one of
the most essential aspects of the process of listening to music.
A veridical computational model of perceived musical tension would
be an important ingredient for many music informatics applications
\cite{manifesto}.
The present paper presents a new approach
to modelling harmonic tension based on a distributed representation of chords.
The starting hypothesis is that harmonic tension as perceived by
human listeners is related, among other things, to the expectedness of
harmonic units (chords) in their local harmonic context. We train a
\textit{word2vec}-type neural network to learn a vector space that
captures contextual similarity and expectedness, and define a quantitative
measure of harmonic tension on top of this. To assess the veridicality
of the  model, we compare its outputs on a number of well-defined chord
classes and cadential contexts to results from pertinent empirical studies
in music psychology. Statistical analysis shows that the model's
predictions conform very well with empirical evidence obtained
from human listeners.

\end{abstract}

\section{Introduction}\label{sec:introduction}
Musical  tension results from the action and interaction among numerous musical features, such as the gradual rise in loudness or pitch height at the climax of a Galant symphony (the so-called \textit{Mannheim crescendo}), or the increase in tempo, rhythmic variability, or onset density in the development section of a classical sonata-form movement \cite{Farbood2012,illia2011}. Indeed, even an isolated repeating tone develops a different quality and different impact on listeners after each repetition \cite{Margulis2016}, thus awakening the potential for tension during music listening. However, in this study we will restrict the discussion to the \textit{harmonic} tension resulting from simultaneously-sounding tones in a polyphonic (i.e., multi-voiced) sequence and the context in which these appear.

Among music theorists, harmonic tension is often characterized as the sense that an unstable harmonic state tends to resolve to another more calm (or relaxed) state. The resolution of a chord can occur immediately from one chord to the next, or it may be delayed via techniques of extension and prolongation, such as the pedal or organ point at the end of a Baroque fugue \cite{Meyer1973}.

Psychologists have generally offered either sensory (or psychoacoustic) or cognitive explanations for the perception of harmonic tension. On the one hand, sensory accounts characterize harmonic tension as a sensory response caused by (1) the rapid beating (oscillations in amplitude) created by interactions of adjacent partials, and (2) the absence of shared partials between two or more complex tones (called \textit{inharmonicity}) \cite{Helmholtz1865}. On the other hand, cognitive accounts treat tension as an emotional affect resulting from the formation, fulfilment, and violation of expectations learned from exposure to Western music \cite{Meyer1956}.

This study offers a cognitive account for the experience of harmonic tension using \textit{word2vec}, a neural network-based representation model that simulates the long-term learning of words in natural languages. Our approach is mainly motivated by two assumptions: (1) that chords---like words in a natural language corpus---are similar to each other if they occur in similar (or identical) contexts, and (2) that the ebb and flow of tension/relaxation is caused by changes in the (dis)similarity between adjacent chords. These assumptions are consistent with experimental studies in music perception and cognition that underline the role of harmonic context in the formation of expectations during music listening \cite{Bigand1997,BigandMadurell1999,Schmuckler1994,TillmannBigand2001}, in which chords that co-occur frequently in a surrounding context were found to be more expected by listeners.

We begin in Section 2 by reviewing chord typologies in music theory and the related evidence for harmonic tension based on sensory and cognitive accounts. Next, Section 3 presents a new model of harmonic tension based on word2vec and describes the datasets used in the present research. Finally, the remaining two sections present the results of two experiments designed to replicate the findings from previous studies with human participants using the word2vec model; Section 4 examines harmonic tension for isolated sonorities like triads and seventh chords, while Section 5 considers the dynamic variations of tension accompanying common chord progressions in classical music (called \textit{cadences}). 

\section{Harmonic tension in chords and chord sequences}
\begin{figure}[t!]
	\centering
	\includegraphics[width=12cm]{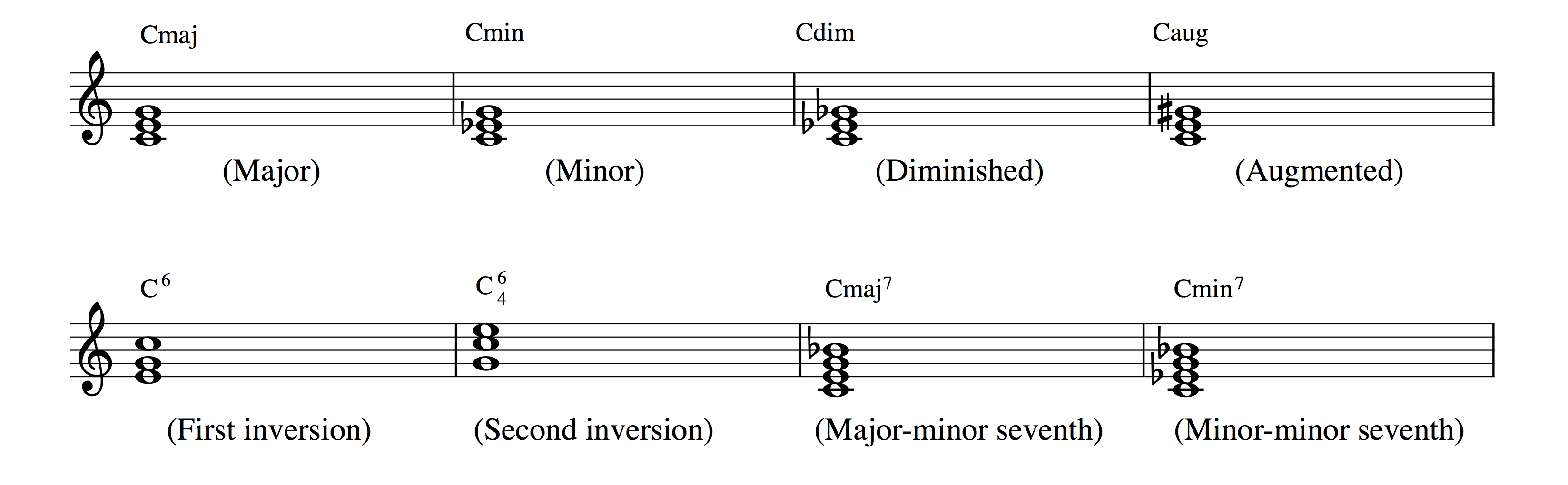}
	\caption{Examples of different chord qualities, inversions and types (all notated with C roots). }
	\label{chords}
\end{figure}

A chord is a combination of at least two simultaneously sounding tones. This paper considers two types of chords: triads and seventh chords. Depending on the interval structure, various chord qualities emerge (e.g. major, minor, diminished, augmented). Furthermore, a chord is inverted if any pitch other than the root is in the lowest (bass) voice (see Figure \ref{chords}). We will call the units of time over which these chords occur \textit{harmonic units}.

As mentioned before, sensory models of tension investigate the effect of roughness or beating on listeners' perception of the dissonance of chords. For instance, \cite{Hutchinson1979} presented a sensory model that calculates the roughness of isolated three- or four-note chords and reported high dissonance estimates for chords that contain dissonant intervals (e.g., seconds or sevenths) and relatively low dissonance estimates for major and minor triads and their inversions. Another model in \cite{Parncutt1989} computed the strength or clarity (vs. ambiguity) of the root pitch of a chord under the assumption that chords with a clear root have a strong harmonic function and are therefore more consonant. The results were consistent with music-theoretical assumptions and empirical studies \cite{Cook2006,Laird2012}. However, these models delivered a rank ordering from the most consonant to the most dissonant triad quality (major $>$ minor $>$ augmented $>$ diminished) that is only partly consistent with the responses of listeners in experimental tasks (major $>$ minor $>$ diminished $>$ augmented) \cite{Arthurs2013,Cook2006,Laird2012,Roberts1986}. It seems that the ranks associated with diminished and augmented triads cannot be calculated by existing sensory models \cite{Cook2006}.

The effect of the musical context in which a chord occurs on listeners' perception of dissonance and consonance has been considered in \cite{Laird2012}, which showed that major and minor triads tend to be more consonant if they occur in a compatible tonal context. Furthermore, listeners judge diminished chords as less dissonant if they occur on the fourth (or \textit{subdominant}) degree of the diatonic scale, and augmented chords as most dissonant if they occur on the first (or \textit{tonic}) scale degree \cite{Arthurs2013}. What is more, both diminished and augmented triads are less tense when they occur as part of a \textit{cadence} \cite{Arthurs2013}, a melodic-harmonic formula that marks the conclusion of a musical phrase \cite{Caplin1998}.

Previous studies have underlined the role of the harmonic functions of chords in a musical context for listeners' expectations \cite{Bigand1997,BigandMadurell1999,TillmannBigand2001}, demonstrating that harmonic units with a strong harmonic relationship to the musical context are more expected. In \cite{Sears2016}, for example, cadences with a tonic terminal chord were found to be more expected than cadences with a non-tonic chord.
Furthermore, \cite{Schmuckler1994} reported that listeners respond faster and more accurately to a chord at the end of a sequence if it belongs to the musical context.

Taken together, these studies suggest that harmonic tension is highly related to the musical context in which these chords occur. Chords that belong to a given harmonic context are more expected, and are therefore less tense. Thus, the present study examines whether an unsupervised learning algorithm can simulate the behavior of listeners for both isolated chords and chord progressions.

\section{Modeling tension with word2vec}
To provide tension estimates for every harmonic unit in the corpus, we employ the word2vec algorithm developed in \cite{Mikolov2013b,Mikolov2013}, a neural network model that creates a distributed representation of words from natural language corpora. The word2vec algorithm has been used as part of the natural language processing pipeline for many state-of-the-art deep learning models to reduce the dimensionality of word vocabularies, but also in other machine learning domains involving symbolic music, such as the word2vec-based chord recommendation system in \cite{Huang2016}.

A formal description of word2vec can be found in \cite{Mikolov2013b,Mikolov2013}. In short, word2vec calculates the similarity between words based on the frequency of their co-occurrence in a surrounding context. For instance the word \textit{like} will be similar to \textit{love}, since \textit{like} and \textit{love} occur frequently in the same surrounding context (e.g., I like this book vs. I love this book). This co-occurrence matrix is then used to create an embedding space of words, where each word is represented as a vector, and where the similarity between words is calculated as the cosine similarity of their vectors. In such spaces, words that appear in similar contexts therefore feature shorter cosine distances (e.g., \textit{small} and \textit{smaller}) compared to those in dissimilar contexts (e.g., \textit{small} and \textit{France}) \cite{Mikolov2013b}.

From a music-theoretical point of view, the idea of the similarity of chords based on their surrounding context corresponds well with functional theories of tonal harmony. In \cite{Riemann}, for example, the harmonic behaviours of chords is reduced to three functional categories---tonic, dominant and predominant. The harmonic function of a given chord therefore depends on the tonal context in which it occurs. Thus, altering a chord likely does not change its harmonic function, and so surrounding context similarity can be interpreted as similarity of harmonic function. 

As an example, consider a very common chord progression in C-major: \textit{C -- F -- G -- C} (i.e., \textit{I -- IV -- V -- I}). The surrounding context of the dominant chord \textit{G} includes \textit{C} and \textit{F}. In practice, we might observe some alterations of the dominant chord, like \textit{G\textsuperscript{7}}. Yet despite these alterations, its harmonic function remains the same: namely, dominant. In other words, in the context of C-major, \textit{G} and \textit{G\textsuperscript{7}} are very similar to each other (even though they would be highly dissimilar in the context of G-major).  What is more, if a new chord is inserted at the end of the sequence above, \textit{G} will be less tense than \textit{Am}, since our memory already contains the harmonic function of dominant (\textit{G}) but not the function associated with the submediant (\textit{Am}). 
\subsection{The Vocabulary}
The word2vec algorithm requires a sequence of symbols as input, so we represent each harmonic unit as an integer ID in the model. To extract the harmonic units from each piece, we first produce a \textit{full expansion} of the symbolic encoding \cite{Conklin2002}, which duplicates overlapping note events at every unique onset time. To decrease the size of the vocabulary, the pitches in each harmonic unit are then mapped as a set of \textit{pitch classes}, which reduces the domain of possible pitches to the twelve-tone chromatic scale (e.g., C$=0$, C$\sharp$/D$\flat=1$, etc.). Finally, duplicated pitch classes are removed.

\subsection{The Corpus}
The corpus consists of 297 string quartet movements composed during the high classical period by Josef Haydn (215; 1762--1803) and Wolfgang Amadeus Mozart (82; 1770--1790). All movements were downloaded from the KernScores database in kern format.\footnote{\url{http://kern.ccarh.org/}.} To create a dataset that equally represents all possible keys from the twelve-tone chromatic scale, all of the movements were transposed 11 times from $-5$ to 6 semitones. This procedure produced a vocabulary consisting of 4753 possible chord types in the corpus. The sequence of harmonic units from each piece was then converted into a sequence of integer IDs corresponding to the chord types from the vocabulary.

\subsection{Training the model}
The original implementation of word2vec was used to generate a distributed representation of harmonic units.\footnote{\url{https://code.google.com/p/word2vec/}.} The model was trained with the Continuous Bag of Words (or $cbow$) option, which predicts each word given the words in the surrounding context. The number of dimensions of the embedding space was set to 120 and the threshold for the required minimum number of word occurrences was set to 1 to consider all words. Finally, the model was trained with a window size of 6 harmonic units.

\subsection{Calculating tension}
We define the tension estimate of a harmonic unit $H_t$ as the average cosine distance between $H_t$ and $n$ preceding harmonic units, which was set to 24 for the experiments reported here. To simulate the decay in memory over time, each of the preceding units is multiplied by a weighting function $1-e^{1-\frac{1}{(i-1)/n}}$, which ensures that the cosine distances estimated from recently heard harmonic units receive higher weights than those from more remote harmonic units in the sequence. 
\begin{equation}
tension(H_t) = -\frac{1}{n}\sum_{i=1}^n
cos({H_{t-i}},H_t)(1-e^{1-\frac{1}{(i-1)/n}})
\end{equation}

The output of cosine distance lies on a scale between $-1$ and 1, where the largest value 1 indicates minimum distance (i.e., minimum tension in our model). However, to improve the legibility of the output, we multiply the weighted output of our function with $-1$ so that lower values indicate less tension, and a value of 0 indicates maximum tension.

\section{Experiment 1}
Experiment 1 examines the average tension estimates for different chord categories according to three chord conditions: \textit{type} (triad, seventh), \textit{quality} (major, minor, diminished, augmented), and \textit{inversion} (root, first, second, third). Major and minor triads in root position are generally expected to be less tense than those in first or second inversion \cite{Cook2006}. In addition, the following trend from the most consonant to most dissonant triad quality is assumed; major, minor, diminished and augmented \cite{Cook2006,Laird2012,Roberts1986}. Furthermore, seventh chords are assumed to be more tense than triads.

\subsection{Analysis}
To evaluate the tension estimates for all chord categories in the corpus, we applied 10-fold cross validation, in which the corpus was divided into 10 subsets and trained 10 times, each time leaving out a different subset for testing. Each condition in this experiment refers to a chord category which is defined by three chord aspects: type, quality and inversion. Due to large differences in sample size across chord conditions, the number of chords in each condition was limited to 1200 chords selected randomly from the corpus.

To examine the tension estimates of all categories, we considered the following four hypotheses: 

\begin{enumerate}[label=\subscript{H}{\arabic*}]
	\item Major triads in first inversion are more tense than those in root position or in second inversion \cite{Hutchinson1979}.
	\item Minor triads in root position are less tense than those in first or second inversion \cite{Roberts1986}.
	\item Regarding chord quality, the following rank for increasing tension obtains: major $<$ minor $<$ diminished $<$ augmented \cite{Cook2006,Laird2012,Roberts1986}.
	\item Triads are less tense than seventh chords.
\end{enumerate}

Levene's test indicated unequal variances for every between-groups factor for chord categories, so we report Type III ANOVAs with White correction \cite{White1980}, which is used to compute heteroscedasticity-robust covariance matrices. To examine differences between the levels of each factor, we also included four planned comparisons that do not assume homogeneity of variances. The first two comparisons examine the potential differences between the levels of inversion for major (H$_1$) and minor (H$_2$) triads. The next comparison examines the predicted trend across the levels of quality (H$_3$). Finally, the fourth comparison considers the potential difference between triads and seventh chords (H$_4$). To minimize the chance of making a Type I error, we also apply a Bonferroni correction to the planned comparisons, which divides the critical \textit{p} value by the number of comparisons ($\frac{.05}{8}$, so alpha \textsubscript{crit}$= .00625$). As a consequence, only \textit{p} values smaller than .00625 will be significant.

\subsection{Results}

To examine the first two hypotheses, tension estimates for all triad categories were submitted to a two-way ANOVA with factors of \textit{quality} (major, minor, diminished, augmented) and \textit{inversion} (root, first, second). The interaction between quality and inversion was significant, $F(4, 10556) = 35.8, p < .001,  \eta_{p}^{2} = .01)$, so we examined simple main effects of quality and inversion for the first two hypotheses.

Figure \ref{major-minor-Inversions} presents the average tension estimates for major and minor triads for each level of inversion. Beginning with H$_1$, a one-way ANOVA revealed a simple main effect of inversion for major triads, $F(2, 3558) = 76.56, p < .001,  \eta_{p}^{2} = .04$. As expected, triads in first inversion elicited higher tension estimates than those either in root position, $t(3558) = -12.24, p < .001$, or in second inversion, $t(3558) = -7.56, p < .001$.

\newlength\figureheight
\newlength\figurewidth
\setlength\figureheight{10cm}
\setlength\figurewidth{6cm}

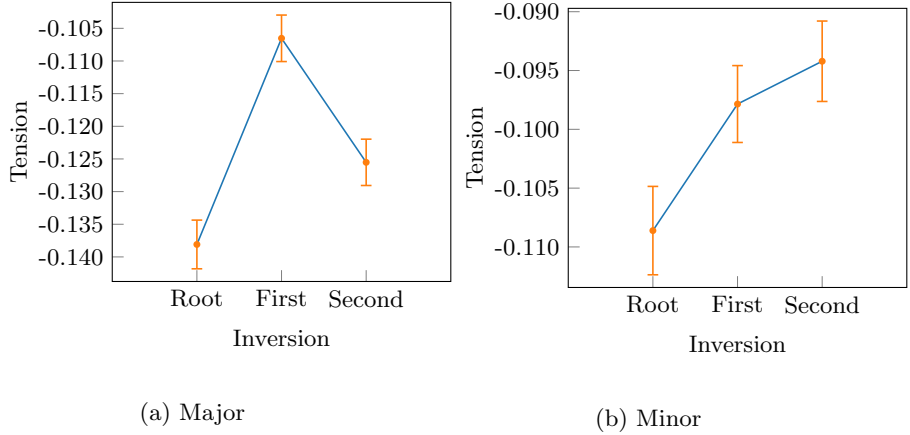
\begin{figure}[t!]

\begin{subfigure}{5cm}

 \centering
  \setlength\figureheight{\linewidth} 
  \setlength\figurewidth{\linewidth}
\begin{tikzpicture}

\definecolor{color1}{rgb}{0.12156862745098,0.466666666666667,0.705882352941177}
\definecolor{color0}{rgb}{1,0.498039215686275,0.0549019607843137}

\begin{axis}[
scale=0.65,
xlabel={Inversion},
ylabel={Tension},
xmin=-1, xmax=3,
ymin=-0.143761297720379, ymax=-0.101025570130305,
xtick={0,1,2},
xticklabels={Root,First,Second},
ytick={-0.145,-0.140,-0.135,-0.130,-0.125,-0.120,-0.115,-0.110,-0.105,-0.10},
yticklabels={,-ˆ'0.140,-0.135,-ˆ'0.130,-ˆ'0.125,-ˆ'0.120,-ˆ'0.115,-ˆ'0.110,-ˆ'0.105,},
tick align=inside,
tick pos=left,
x grid style={lightgray!92.026143790849673!black},
y grid style={lightgray!92.026143790849673!black}
]
\path [draw=color0, semithick] (axis cs:0,-0.141818764648103)
--(axis cs:0,-0.134363952175564);

\path [draw=color0, semithick] (axis cs:1,-0.110086960057752)
--(axis cs:1,-0.102968103202581);

\path [draw=color0, semithick] (axis cs:2,-0.129064522614961)
--(axis cs:2,-0.121971848122334);

\addplot [semithick, color1]
table {%
0 -0.138091358411833
1 -0.106527531630167
2 -0.125518185368648
};
\addplot [semithick, color0, mark=-, mark size=2, mark options={solid}, only marks]
table {%
0 -0.141818764648103
1 -0.110086960057752
2 -0.129064522614961
};
\addplot [semithick, color0, mark=-, mark size=2, mark options={solid}, only marks]
table {%
0 -0.134363952175564
1 -0.102968103202581
2 -0.121971848122334
};
\addplot [semithick, color0, mark=*, mark size=1, mark options={solid}, only marks]
table {%
0 -0.138091358411833
1 -0.106527531630167
2 -0.125518185368648
};

\end{axis}

\end{tikzpicture} \caption{Major}
\end{subfigure}%
\hspace{1cm}%
\begin{subfigure}{5cm}
\begin{tikzpicture}

\definecolor{color1}{rgb}{0.12156862745098,0.466666666666667,0.705882352941177}
\definecolor{color0}{rgb}{1,0.498039215686275,0.0549019607843137}

\begin{axis}[
scale=0.65,
xlabel={Inversion},
ylabel={Tension},
xmin=-1, xmax=3,
ymin=-0.113447667543031, ymax=-0.089715141990818,
xtick={0,1,2},
xticklabels={Root,First,Second},
ytick={-0.115,-0.110,-0.105,-0.100,-0.095,-0.090,-0.085},
yticklabels={,-ˆ'0.110,-ˆ'0.105,ˆ'-0.100,-ˆ'0.095,-ˆ'0.090,},
tick align=inside,
tick pos=left,
x grid style={lightgray!92.026143790849673!black},
y grid style={lightgray!92.026143790849673!black}
]
\path [draw=color0, semithick] (axis cs:0,-0.112368916381567)
--(axis cs:0,-0.104854108987561);

\path [draw=color0, semithick] (axis cs:1,-0.101118321471563)
--(axis cs:1,-0.0945819574486829);

\path [draw=color0, semithick] (axis cs:2,-0.0976309047929957)
--(axis cs:2,-0.0907938931522823);

\addplot [semithick, color1]
table {%
0 -0.108611512684564
1 -0.0978501394601227
2 -0.094212398972639
};
\addplot [semithick, color0, mark=-, mark size=2, mark options={solid}, only marks]
table {%
0 -0.112368916381567
1 -0.101118321471563
2 -0.0976309047929957
};
\addplot [semithick, color0, mark=-, mark size=2, mark options={solid}, only marks]
table {%
0 -0.104854108987561
1 -0.0945819574486829
2 -0.0907938931522823
};
\addplot [semithick, color0, mark=*, mark size=1, mark options={solid}, only marks]
table {%
0 -0.108611512684564
1 -0.0978501394601227
2 -0.094212398972639
};

\end{axis}

\end{tikzpicture} \caption{Minor}
\end{subfigure}
\caption{Mean tension estimates for major and minor triads in root position, and in first and second inversion. Whiskers represent $\pm$2 standard errors.}%
	\label{major-minor-Inversions}
\end{figure}

For H$_2$, a one-way ANOVA carried out on minor triads also revealed a simple main effect of inversion $F(2, 3463) = 17.02, p < .001,  \eta_{p}^{2} = 0.01$. In this case, triads in root position elicited significantly lower tension estimates than those either in first inversion, $t(3463) = 4.32, p < .001$, or in second inversion $t(3463) = 5.67, p < .001$.

\begin{figure}[t!]
\begin{subfigure}{5cm}
 \centering
  \setlength\figureheight{\linewidth} 
  \setlength\figurewidth{\linewidth}
\begin{tikzpicture}

\definecolor{color1}{rgb}{0.12156862745098,0.466666666666667,0.705882352941177}
\definecolor{color0}{rgb}{1,0.498039215686275,0.0549019607843137}

\begin{axis}[
scale=0.65,
xlabel={Quality},
ylabel={Tension},
xmin=-1, xmax=4,
ymin=-0.129517020208519, ymax=-0.0408520969315872,
xtick={0,1,2,3},
xticklabels={Maj.,Min.,Dim.,Aug.},
ytick={-0.13,-0.12,-0.11,-0.10,-0.09,-0.08,-0.07,-0.06,-0.05,-0.04},
yticklabels={,-ˆ'0.12,-ˆ'0.11,-ˆ'0.10,-ˆ'0.09,-ˆ'0.08,-ˆ'0.07,-ˆ'0.06,-ˆ'0.05,},
tick align=inside,
tick pos=left,
x grid style={lightgray!92.026143790849673!black},
y grid style={lightgray!92.026143790849673!black}
]
\path [draw=color0, semithick] (axis cs:0,-0.125486796423204)
--(axis cs:0,-0.121224397771348);

\path [draw=color0, semithick] (axis cs:1,-0.102391114189029)
--(axis cs:1,-0.0983328213447279);

\path [draw=color0, semithick] (axis cs:2,-0.0698712242575283)
--(axis cs:2,-0.0662905464697753);

\path [draw=color0, semithick] (axis cs:3,-0.0504079755353101)
--(axis cs:3,-0.0448823207169023);

\addplot [semithick, color1]
table {%
0 -0.123355597097276
1 -0.100361967766878
2 -0.0680808853636518
3 -0.0476451481261062
};
\addplot [semithick, color0, mark=-, mark size=2, mark options={solid}, only marks]
table {%
0 -0.125486796423204
1 -0.102391114189029
2 -0.0698712242575283
3 -0.0504079755353101
};
\addplot [semithick, color0, mark=-, mark size=2, mark options={solid}, only marks]
table {%
0 -0.121224397771348
1 -0.0983328213447279
2 -0.0662905464697753
3 -0.0448823207169023
};
\addplot [semithick, color0, mark=*, mark size=1, mark options={solid}, only marks]
table {%
0 -0.123355597097276
1 -0.100361967766878
2 -0.0680808853636518
3 -0.0476451481261062
};

\end{axis}

\end{tikzpicture} \caption{Triads}
  \label{chordordering}
\end{subfigure}%
\hspace{1cm}%
\begin{subfigure}{5cm}
 \centering
  \setlength\figureheight{\linewidth} 
  \setlength\figurewidth{\linewidth}

\begin{tikzpicture}

\definecolor{color1}{rgb}{0.12156862745098,0.466666666666667,0.705882352941177}
\definecolor{color0}{rgb}{1,0.498039215686275,0.0549019607843137}

\begin{axis}[
scale=0.65,
xlabel={Quality},
ylabel={Tension},
xmin=-1, xmax=2,
ymin=-0.115064620791649, ymax=-0.0809046411935604,
xtick={0,1},
xticklabels={Triad,Seventh},
ytick={-0.120,-0.115,-0.110,-0.105,-0.100,-0.095,-0.090,-0.085,-0.080},
yticklabels={,-0.115,-0.110,-ˆ'0.105,-ˆ'0.100,-ˆ'0.095,-0.090,-ˆ'0.085,},
tick align=inside,
tick pos=left,
x grid style={lightgray!92.026143790849673!black},
y grid style={lightgray!92.026143790849673!black}
]
\path [draw=color0, semithick] (axis cs:0,-0.113511894446281)
--(axis cs:0,-0.110516527794618);

\path [draw=color0, semithick] (axis cs:1,-0.0846970934117046)
--(axis cs:1,-0.082457367538928);

\addplot [semithick, color1]
table {%
0 -0.11201421112045
1 -0.0835772304753163
};
\addplot [semithick, color0, mark=-, mark size=2, mark options={solid}, only marks]
table {%
0 -0.113511894446281
1 -0.0846970934117046
};
\addplot [semithick, color0, mark=-, mark size=2, mark options={solid}, only marks]
table {%
0 -0.110516527794618
1 -0.082457367538928
};
\addplot [semithick, color0, mark=*, mark size=1, mark options={solid}, only marks]
table {%
0 -0.11201421112045
1 -0.0835772304753163
};

\end{axis}

\end{tikzpicture} \caption{Triads and seventh chords}
   \label{triadSeventh}
\end{subfigure}
\caption{Mean tension estimates for (a) major, minor, diminished and augmented triads and for (b) triads, and seventh chords. Whiskers represent $\pm$2 standard errors.}%
\end{figure}
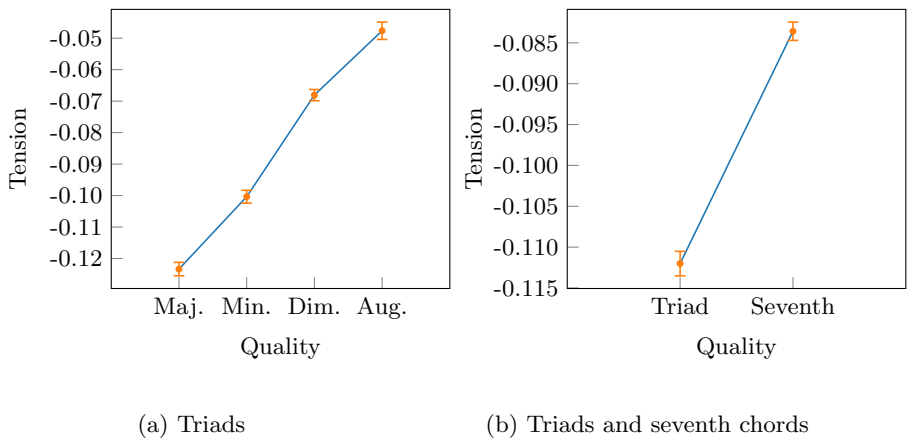

Figure \ref{chordordering} presents the average tension estimates for all triads. A reverse forward-difference contrast revealed that the mean tension estimates increased significantly for major vs. minor triads, $t(11465) = -41.2, p < .001$, minor vs. diminished triads, $t(11465) = -48.9, p < .001$, and diminished vs. augmented triads, $t(11465) = -33.15, p < .001$. Thus, the tension estimates revealed a clear ascending linear trend from the major to the augmented triad conditions, thereby demonstrating the predicted trend in H$_3$.

\section{Experiment 2}
In this experiment, we examined the average tension estimates for the terminal chords from cadences. In \cite{Sears2016}, a collection of cadences was identified in a corpus of 50 Haydn string quartets (Opp. 17--76) using Caplin's cadence typology \cite{Caplin1998,Caplin2004}, so we selected the three most common cadence categories from that collection. The perfect authentic cadence (PAC) is the strongest cadence type, and features a chord progression from a root-position dominant to a root-position tonic (in C-major, \textit{G} to \textit{C}). Next, the half cadence (HC) terminates with dominant harmony, and is generally assumed to be less stable, and thus more tense. Finally, the deceptive cadence (DC) deceives listeners by initially promising a PAC, but then arriving at a non-tonic harmony, usually vi (in C-major, \textit{G} to \textit{Am}). The cadence collection consists of 122 PACs, 84, HCs, and 19 DCs, for a total of 225 example cadences. 

To examine how the model calculates tension estimates for these cadence categories, we considered the following four hypotheses:

\begin{enumerate}[label=\subscript{H}{\arabic*}]
	\item Terminal chords from PACs elicit lower tension estimates than those from non-cadential chords selected at random from the corpus ($PAC <$ \textit{non-cad}).
	\item Terminal chords from HCs are more tense than those from PACs because they end with the less stable dominant harmony ($PAC < HC$).
	\item Terminal chords from DCs are more tense than those from PACs because they end with an unexpected non-tonic harmony ($PAC < DC$). 
	\item The following rank for increasing tension obtains: $PAC < HC < DC$.
\end{enumerate}

\subsection{Analysis}
To evaluate model performance, we again used the Haydn and Mozart string quartets corpus for training. However, since annotated movements from Haydn's cadence collection also appear in the training set, we applied 10-fold cross validation and randomly divided movements containing the cadence collection into 10 equal disjunct groups.

\subsection{Results}

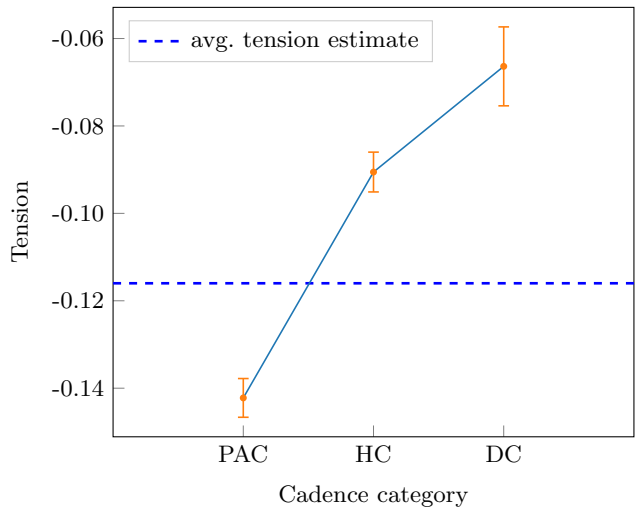
\begin{figure}[t!]
\centering
\begin{tikzpicture}

\definecolor{color1}{rgb}{0.12156862745098,0.466666666666667,0.705882352941177}
\definecolor{color0}{rgb}{1,0.498039215686275,0.0549019607843137}

\begin{axis}[
xlabel={Cadence category},
ylabel={Tension},
xmin=-1, xmax=3,
ymin=-0.151119035224814, ymax=-0.052871305367868,
xtick={0,1,2},
xticklabels={PAC,HC,DC},
yticklabels={,-0.14,-0.12,-ˆ'0.10,-ˆ'0.08,-ˆ'0.06,-0.4,-ˆ'0.2,},
tick align=inside,
tick pos=left,
x grid style={lightgray!92.026143790849673!black},
y grid style={lightgray!92.026143790849673!black},
legend entries={{avg. tension estimate}},
legend cell align={left},
legend style={at={(0.03,0.97)}, anchor=north west, draw=white!80.0!black}
]
\path [draw=color0, semithick] (axis cs:0,-0.146653229322226)
--(axis cs:0,-0.137790819858102);

\path [draw=color0, semithick] (axis cs:1,-0.0950981023784741)
--(axis cs:1,-0.085991366764383);

\path [draw=color0, semithick] (axis cs:2,-0.0753967171505962)
--(axis cs:2,-0.0573371112704565);
\addplot [line width=1.0pt, blue, dashed]
table {%
-1 -0.115995096508333
3 -0.115995096508333
};

\addplot [semithick, color1]
table {%
0 -0.142222024590164
1 -0.0905447345714286
2 -0.0663669142105263
};
\addplot [semithick, color0, mark=-, mark size=2, mark options={solid}, only marks]
table {%
0 -0.146653229322226
1 -0.0950981023784741
2 -0.0753967171505962
};
\addplot [semithick, color0, mark=-, mark size=2, mark options={solid}, only marks]
table {%
0 -0.137790819858102
1 -0.085991366764383
2 -0.0573371112704565
};

\addplot [semithick, color0, mark=*, mark size=1, mark options={solid}, only marks]
table {%
0 -0.142222024590164
1 -0.0905447345714286
2 -0.0663669142105263
};

\end{axis}

\end{tikzpicture}

\caption{Mean tension estimates for the terminal chord of the cadence categories.  Whiskers represent $\pm$1 standard error.}
\label{cadences}
\end{figure}

Figure \ref{cadences} presents the average tension estimates for the terminal chords from all three cadence categories. To calculate the average tension estimate over all pieces (shown with the dotted blue line), we randomly selected 1200 chords from the corpus. For H$_1$ ($PAC < $ \textit{non-cad}), Levene's test indicated unequal variances, so we again report Bonferroni-corrected $t$-tests that do not assume equal variances. In this case, the terminal chords from the PAC category elicited significantly lower tension estimates than the non-cadencial chords that were selected randomly from the copus: $t(1320) = 5.36, p < .001$.

For H$_2$ ($PAC < HC$) and H$_3$ ($PAC < DC$), a one-way ANOVA revealed significant differences between cadence categories, $F(2,222) = 44.43, p < .001, \eta_{p}^{2}  = .28$. The PAC category also elicited significantly lower tension estimates than either the HC category, $t(204) = 7.89, p < .001$, or the DC category, $t(139) = 6.43, p < .001$.

Finally, the reverse forward-difference contrast to examine H$_4$ revealed a significant ascending linear trend from the PAC to the DC categories: (PAC vs. HC, $t(243) = -8.95, p < .001$; HC vs. DC $t(243) = -4.56, p < .001$), thereby demonstrating the predicted trend, $PAC < HC < DC$.

\section{Discussion and Conclusion}
This paper presented a new approach to the calculation of tension estimates of harmonic units based on a distributed representation of chords using word2vec. To obtain the reported tension estimates without requiring subjective judgments from human listeners, we conducted two computational experiments using a corpus of string quartets from the classical style. In the first experiment, word2vec provided tension estimates for different chord categories based on type (triad, seventh), quality (major, minor, diminished, augmented), and inversion (root, first, second). The results corresponded closely with previous studies in music theory and experimental psychology \cite{Arthurs2013,Cook2006,Hutchinson1979,Laird2012,Roberts1986}, demonstrating that an unsupervised learning algorithm can simulate the behavior of listeners in tasks related to the experience of harmonic tension. 

However, the first experiment did not consider the functional context in which these chords occur. Therefore, the second experiment considered the dynamic ebb and flow of tension in tonal harmonic progressions by examining tension estimates for the terminal events from the three most common cadence categories in classical music. The model indicated that the ends of phrases (represented by terminal events in the PAC category) are less tense than non-cadential chord events selected at random from the corpus. Furthermore, the model delivered consistent results with previous research \cite{Sears2016}, which demonstrated that the terminal events from the HC and DC categories are less expected than those from the PAC category. 

To examine how the preceding context influences the experience of harmonic tension, the present model included an exponential decay function that privileges recent harmonic units over temporally remote ones. Over the course of our experiments, we found that as the size of the context increases, the average distance between each cadence category also increases, up to a window size of around 24 units. This finding suggests that the size of the preceding context plays a fundamental role in the experience of harmonic tension, but future studies could examine more carefully how varying the size of the context improves model performance, or define other weighting functions that might better simulate the effects of echoic or short-term memory. In our view, it seems reasonable to suggest that the optimal window size depends on the size and complexity of the harmonic vocabulary: compositions featuring relatively small vocabularies---and in which the number of possible alternatives for each event in the sequence is generally quite low---will require a shorter context size than those featuring larger vocabularies. What is more, given the window size selected here, our results might also reflect a kind of global goodness-of-fit measure between the chords in the sequence and the key in which they occur (i.e., tonal tension).

Since word2vec is a probabilistic model that calculates the similarity between words based on their co-occurrence probability, the tension estimates reported here amount to probabilistic inferences about the co-occurrence of chords in a given harmonic context. As a consequence, this study operationalizes tension as an affective (or emotional) response to the formation, fulfillment, and violation of expectations during music listening \cite{Meyer1956,Meyer1973}. However, this is not to say that tension cannot refer to other aspects of musical experience (e.g. melodic organization), or result from other (sensory or psychoacoustic) mechanisms during perception (e.g., from sensory dissonance, inharmonicity, or auditory roughness). Rather, this study simply reinforces the view that unsupervised learning algorithms can induce the statistical regularities governing complex stimulus domains like natural language and tonal music, and that aspects of emotional experience---like tonal (or harmonic) tension---result from this learning process.


\end{document}